\documentclass[twocolumn,showpacs,prl]{revtex4}%
\usepackage{amsfonts}
\usepackage{amsmath}
\usepackage{amssymb}
\usepackage{graphicx}
\begin{document}

\title{Decoherence of coupled electron spins via nuclear spin dynamics in quantum dots}

\author{W. Yang}
\affiliation{Department of Physics, The Chinese University of Hong Kong, Shatin, N. T., Hong Kong, China}
\author{R. B. Liu}
\thanks{Email: rbliu@phy.cuhk.edu.hk}
\affiliation{Department of Physics, The Chinese University of Hong Kong, Shatin, N. T., Hong Kong, China}


\begin{abstract}
In double quantum dots, the exchange interaction between two electron spins
renormalizes the excitation energy of pair-flips in the nuclear spin bath, which
in turn modifies the non-Markovian bath dynamics. As the energy renormalization varies
with the Overhauser field mismatch between the quantum dots, the electron singlet-triplet
decoherence resulting from the bath dynamics depends on sampling of nuclear spin
states from an ensemble, leading to the transition from exponential decoherence in
single-sample dynamics to power-law decay under ensemble averaging. In contrast,
the decoherence of a single electron spin in one dot is essentially the same
for different choices of the nuclear spin configuration.
\end{abstract}
\maketitle

Decoherence draws a boundary between the microscopic quantum world and the macroscopic classical
world. It is also a main obstacle in quantum technologies such as quantum computation.
Thus, both for understanding crossover from the quantum to
the classical world~\cite{Zurek_decoherence_RMP,Decoherence_classicalWorld,Schlosshauer_decoherence},
and for exploiting quantum coherence of large systems~\cite{Aharonov:2006},
it is desirable to comprehend how decoherence develops with scaling up the
size of a system. The very initial step toward such a purpose is to
examine the difference between a two-level system (the simplest quantum object,
called a qubit in quantum computation) and two coupled ones.
For a system in a Markovian bath (which has broad-band fluctuation),
the decoherence is described by the Lindblad formalism.
For a system in a non-Markovian bath, there are indications of nontrivial scaling behavior of
decoherence~\cite{Burkard:2005,You:2005,Krojanski:2006,Hu_ChargeFluctuation},
such as the non-additive decoherence in multiple baths~\cite{Burkard:2005}.
In this paper, we study the decoherence of a composite quantum object in a non-Markovian mesoscopic
bath, based on a paradigmatic system in mesoscopic physics and quantum information
science~\cite{Loss_QDspinQC,Marcus_T2,Koppens_T2},
namely, two electron spins in double quantum dots (QDs) where the nuclear spins serve as the bath.

The dynamics of a mesoscopic nuclear spin bath in a QD is conditioned on the state of the
electron spin in contact with the bath~\cite{Yao_Decoherence,Yao_DecoherenceControl,Liu_Decoherence}.
The conditional evolution of the nuclear spins establishes the electron-nuclear entanglement
that causes the electron spin decoherence. When the electron spin is disturbed, the
nuclear bath dynamics is altered. For example, the nuclear spin evolution can be shepherded by a structured
sequence of electron spin flips so that the electron is disentangled from the bath and as a result the lost coherence
is recovered~\cite{Yao_DecoherenceControl}. Besides external control, the disturbance may also be
due to interaction with another quantum object in proximity, such as the exchange interaction
$J_{\text{ex}}\hat{\mathbf S}_1\cdot\hat{\mathbf S}_2$ between two electron spins in coupled QDs.
Intuitively speaking, the disturbance due to the exchange interaction may be viewed as precessing
of the electron spins about each other.  For a more rigorous treatment, we should first
diagonalize the electron spin Hamiltonian including the exchange interaction, and then study the bath dynamics
and the decoherence in the electron eigenstate basis. To demonstrate the most essential physics, we consider only
the decoherence between the singlet state $\left|S \right\rangle $ and the unpolarized triplet
state $\left| T_{0}\right\rangle$ (by assuming that the polarized triplet states
$\left| T_{\pm}\right\rangle$ are well split off by a large external magnetic field).
The singlet-triplet (S-T) dephasing due to the Overhauser field distribution in an ensemble or a
superposition state of nuclear spins has been studied in Refs.~\cite{Coish_ST,Taylor:2006}, but
the bath considered there has no dynamical fluctuation since the interaction between nuclear spins
was neglected.

\begin{figure}[t]
\includegraphics[width=\columnwidth]{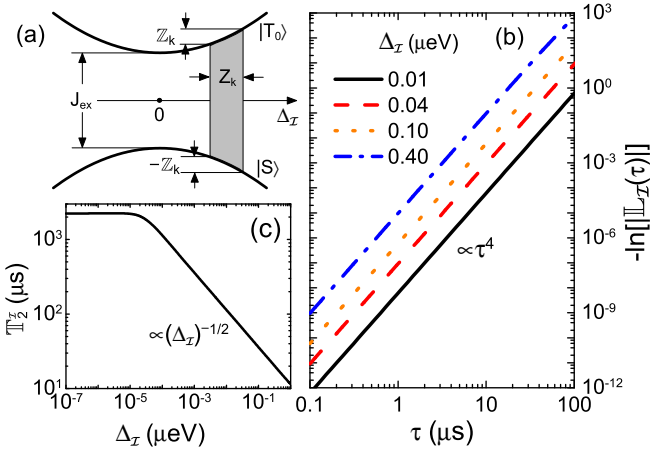}
\caption{(Color online) (a)
The singlet and triplet energies as functions of the Overhauser field mismatch $\Delta_{\mathcal I}$, and the
(renormalized) excitation energy of a nuclear spin pair-flip $Z_k$ (${\mathbb Z}_k$).
(b) FID of the S-T coherence for various initial nuclear spin states, indicated by $\Delta_{\mathcal I}$.
(c) FID decoherence time ${\mathbb T}_{2}^{{\mathcal I}}$ as a function of $\Delta_{\mathcal I}$.}
\label{fig1}
\end{figure}

The role of the exchange interaction in the decoherence may be disclosed in a
semiclassical spectral diffusion picture~\cite{deSousa_Spectral1}. Let us denote the local Overhauser fields for
the nuclear spin configuration $\left|{\mathcal I}\right\rangle$ by the electron Zeeman energies
$\Omega^{\mathcal I}_{1}$ and $\Omega^{\mathcal I}_{2}$ in dot 1 and dot 2, respectively.
The pairwise nuclear spin flip-flops cause the dynamical fluctuation of the Overhauser fields and therefore
a random phase of the electron spins. For a single electron spin in one QD, the Zeeman energy change due to
the $k$th pair-flip is $2{Z}_k$. With exchange interaction, the S-T splitting is
$E_{\text{S-T}}=\sqrt{J^2_{\text {ex}}+\Delta_{\mathcal I}^2}$, varying with the Overhauser field mismatch
$\Delta_{\mathcal I}\equiv \Omega^{\mathcal I}_{1}-\Omega^{\mathcal I}_{2}$ [see Fig.~\ref{fig1}(a)],
So, the S-T splitting change due to the $k$th pair-flip is
\begin{equation}
2{\mathbb Z}_k\approx (2{Z}_k)\frac{\partial E_{\text{S-T}}}{\partial \Delta_{\mathcal I}}\approx  2{Z}_k \frac{\Delta_{\mathcal I}}{E_{\text{S-T}}}.
\label{energycost}
\end{equation}
The exchange interaction modifies the energy fluctuation (or spectral diffusion) responsible for the decoherence.

For a full quantum mechanical description of the
decoherence~\cite{Yao_Decoherence,Yao_DecoherenceControl,Liu_Decoherence,Witzel_Quantum1,Witzel_Quantum2},
we focus on the nuclear spin dynamics which is driven by pair-flips (as elementary excitations in the bath).
The pair-flips are characterized by the flip transition strength and the energy cost.
For uncorrelated electrons in the double dots ($J_{\text{ex}}=0$), the hyperfine-energy cost of a nuclear
pair-flip is $\pm {Z}_k$ for electron spin up and down states, respectively. In presence of exchange
interaction, the energy cost for the pair flip is renormalized to be $\pm {\mathbb Z}_k$ for the triplet and
singlet states, respectively. Thus the bath dynamics itself is altered when non-interacting quantum objects
in the bath are replaced by interacting ones. In particular, through the dependence of the bath fluctuation
on the Overhauser field mismatch $\Delta_{\mathcal I}$, the dynamics of nuclear spins in one dot is
affected by the state of the nuclear spins in the other dot. Therefore, the resultant S-T decoherence
time varies with sampling of the nuclear spin configuration $|{\mathcal I}\rangle$ from an
ensemble $\hat{\rho}_N=\sum_{\mathcal I} P_{\mathcal I}|{\mathcal I}\rangle\langle {\mathcal I}|$.
This, as will be shown later, leads to a transition from exponential decoherence to power-law decay
upon ensemble average. On the contrary, the decoherence of a single electron spin in a QD is essentially
independent of the static Overhauser field~\cite{Yao_Decoherence,Liu_Decoherence}.

We consider a gate-defined double-dot structure similar to those used
in Ref.~\cite{Marcus_T2,Koppens_T2,Coish_ST,Taylor:2006}. We assume a large on-site Coulomb energy and
gate voltages supporting one electron in each dot. The virtual inter-dot tunneling mediates the exchange interaction.
Under a large magnetic field (such that the electron Zeeman energy $\Omega_e\gg J_{\text{ex}}$), the
unpolarized states $\left\vert S\right\rangle$ and $\left\vert T_{0}\right\rangle $
are well separated in energy from the polarized triplet states $\left\vert T_{\pm}\right\rangle$.
The electrons and nuclei are coupled by the contact hyperfine interaction
$\hat{\mathbf S}_1\cdot\hat{\mathbf h}_1+\hat{\mathbf S}_2\cdot\hat{\mathbf h}_2$,
where $\hat{\mathbf h}_{j}=\sum_{n}a_{j,n}\hat{\mathbf I}_{j,n}$, with $a_{j,n}$ being the hyperfine coefficient
for the $n$th nuclear spin at the $j$th dot.
The off-diagonal hyperfine interaction (terms with $\hat{S}^{x}_{j}$, $\hat{S}^y_j$)
couples $|S\rangle$ and $|T_0\rangle$ to $|T_{\pm}\rangle$, which, however,
has negligible effect under a strong external magnetic field ($\Omega_e\ggg a_{j,n}$)~\cite{note_ST_mediated}.
The Overhauser field mismatch $\hat{\Delta}=\hat{h}_{1}^{z}-\hat{h}_{2}^{z}$ couples
$\left\vert T_{0}\right\rangle$ and $\left\vert S\right\rangle$, which causes longitudinal
$T_1$ relaxation. For a relatively large exchange splitting (e.g., $J_{\text{ex}}\gtrsim 10 \Delta_{\mathcal I}$),
the $T_1$ process is suppressed~\cite{Coish_ST,Taylor:2006}, but virtual S-T flips induce a self-energy correction
\begin{equation}
\hat{H}_{\text{S-T}}\equiv\left({E_{\text{S-T}}}+\frac{\Delta_{\mathcal I}}{E_{\text{S-T}}}\hat{\delta}_{\mathcal I}\right)\otimes
\frac{\left|T_{0}\right\rangle \left\langle T_{0}\right|-\left| S\right\rangle \left\langle S\right|}{2},
\end{equation}
where the mean-field part ($\Delta_{\mathcal I}\equiv \langle {\mathcal I}|\hat{\Delta}|{\mathcal I}\rangle$)
is singled out from the quantum fluctuation ($\hat{\delta}_{\mathcal I}\equiv \hat{\Delta}-\Delta_{\mathcal I}$)
of the Overhauser field, and is included non-perturbatively. This self-energy correction is responsible for the
renormalized energy cost (${\mathbb Z}_k$) of a nuclear spin pair-flip driven by the intrinsic
nuclear-nuclear spin interaction $\hat{H}_{N}$.
Now the Hamiltonian of the electron-nuclear spin system is
$
\hat{H}=\hat{H}_+\otimes\left\vert T_{0}\right\rangle \left\langle T_{0}\right\vert
+\hat{H}_-\otimes\left\vert S\right\rangle \left\langle S\right\vert,
$
where $\hat{H}_{\pm}=\hat{H}_{N}\pm E_{\text{S-T}}/2\pm \hat{H}_Z$ with
$\hat{H}_Z=\hat{\delta}_{\mathcal I}\Delta_{\mathcal I}/(2E_{\text{S-T}})$.
Such a block-diagonal Hamiltonian induces no $T_1$ relaxation but only pure S-T dephasing.
We emphasize that the intrinsic nuclear spin interaction $\hat{H}_N$ is essential to
the decoherence. Otherwise, without $\hat{H}_N$, the pure dephasing caused by the
hyperfine interaction is totally eliminated from spin echo
signals~\cite{Yao_Decoherence,Yao_DecoherenceControl,Liu_Decoherence,Shenvi_scaling}.

The S-T decoherence is caused by the electron-nuclear entanglement, established during the evolution of
the nuclear spin state predicated on the electron states:
Suppose the electrons are initially in a superposition state $\alpha|S\rangle+\beta|T_0\rangle$ and the
nuclear state $|{\mathcal I}\rangle$ is one randomly chosen from a thermal ensemble.
The conditional nuclear spin evolution
$|{\mathcal I}\rangle\rightarrow\left|{\mathcal I}^{\pm}({\tau})\right\rangle\equiv e^{-i\hat{H}_{\pm}{\tau}}\left|{\mathcal I}\right\rangle$
establishes an entangled state
$\alpha|S\rangle\otimes|{\mathcal I}^-\rangle+\beta |T_0\rangle\otimes|{\mathcal I}^{+}\rangle$.
The S-T coherence is given by ${\mathbb L}_{\mathcal I}({\tau})=\left\langle {\mathcal I}^-({\tau})\left|{\mathcal I}^+({\tau})\right\rangle\right.$.
To calculate the nuclear spin evolution $|{\mathcal I}^{\pm}({\tau})\rangle$, we employ the pair-correlation approximation
in which all possible pair-flips from the initial configuration $|{\mathcal I}\rangle$ are taken as independent of
each other. The pair-correlation approximation is justified for a mesoscopic nuclear spin bath with a sufficiently
random configuration where within the decoherence timescale the occurred pair-flips are much fewer than the available
pairs to be flipped and therefore have little probability to be in neighborhood of each other or to be
correlated~\cite{Yao_Decoherence,Yao_DecoherenceControl,Liu_Decoherence,Witzel_Quantum1,Witzel_Quantum2,Saiken_Cluster,Witzel:2006}.
A pair-flip is characterized by a transition strength due to the off-diagonal nuclear spin
interaction $B_k=\langle {\mathcal I}|\hat{H}_N|{\mathcal I},k\rangle$,
an energy cost due to the diagonal nuclear spin interaction
$D_k=\langle {\mathcal I},k|\hat{H}_N|{\mathcal I},k\rangle-\langle {\mathcal I}|\hat{H}_N|{\mathcal I}\rangle$,
and a hyperfine energy cost $\pm {\mathbb Z}_k=\pm \left(\langle {\mathcal I},k|\hat{H}_Z|{\mathcal I},k\rangle-\langle {\mathcal I}|\hat{H}_Z|{\mathcal I}\rangle\right)$
for the triplet and the singlet state, respectively, where $|{\mathcal I},k\rangle$ denotes the nuclear spin state
after the $k$th pair-flip. An independent pair-flip is mapped to be a pseudo-spin $\hat{\mathbf s}_k$ precessing about
a pseudo-field ${\boldsymbol \chi}_k^{\pm}\equiv (2B_k,0,D_k\pm {\mathbb Z}_k)$, initially pointing
to the down direction~\cite{Yao_Decoherence,Yao_DecoherenceControl,Liu_Decoherence}.
The nuclear Hamiltonian in the pseudo-spin representation is
$
\hat{H}_{\pm} \approx \pm{E_{\text{S-T}}}/{2}+\sum_{k}{\boldsymbol \chi}^{\pm}_k\cdot\hat{\mathbf s}_{k}.
$

Within the pair-correlation approximation, the S-T coherence in free-induction decay (FID) is
\begin{equation}
{\mathbb L}_{\mathcal I}({\tau})=e^{-iE_{\text{S-T}}{\tau}}\prod_k\left|\left\langle\downarrow\left|e^{+i{\boldsymbol \chi}^-_k\cdot\hat{\mathbf s}_k {\tau}}
e^{-i{\boldsymbol \chi}^+_k\cdot\hat{\mathbf s}_k {\tau}}\right|\downarrow\right\rangle\right|,
\label{FID}
\end{equation}
The short-time behaviour (for ${\tau}\ll {\mathbb Z}_k^{-1}$) is a quartic exponential decay,
${\mathbb L}_{\mathcal I}({\tau})\approx\exp(-iE_{\text{S-T}}{\tau}-\sum_{k}B_{k}^{2}{\mathbb Z}_{k}^{2}{\tau}^{4}/2)
\equiv e^{-iE_{\text{S-T}}{\tau}-({\tau}/{\mathbb T}^{\mathcal I}_2)^4}$.
The decoherence time ${\mathbb T}_{2}^{\mathcal I}\propto \left({\mathbb Z}_k\right)^{-1/2}\propto\left(\Delta_{\mathcal I}/E_{\text{S-T}}\right)^{-1/2}$,
varying with sampling of the nuclear spin configuration from the ensemble $\sum_{\mathcal I}P_{\mathcal I}|{\mathcal I}\rangle\langle {\mathcal I}|$.
For comparison, the decoherence of a single electron spin in a QD~\cite{Note_on_single},
except for a trivial global phase factor related to the inhomogeneous broadening,
is essentially independent of choice of the initial state~\cite{Yao_Decoherence,Yao_DecoherenceControl,Liu_Decoherence},
since the excitation energy of a nuclear pair-flip there
${Z}_k=\left\langle {\mathcal I},k\left|\hat{h}^z_{j}/2\right|{\mathcal I},k\right\rangle -\left\langle {\mathcal I}\left|\hat{h}^z_{j}/2\right|{\mathcal I}\right\rangle$
is independent of $|{\mathcal I}\rangle$. Also, the S-T decoherence time is longer than the single spin decoherence time by
a factor of $\sqrt{E_{\text{S-T}}/\Delta_{\mathcal I}}$ because of the reduction of the excitation energy.

In numerical evaluation, we take a symmetric GaAs double-dot structure with height 6~nm, Fock-Darwin
radius 70~nm for a parabolic confinement potential,
and center-to-center separation 137~nm, under a perpendicular magnetic field $B=1$~T at temperature $T=1$~K.
The calculated variance of the local Overhauser field mismatch is $\Gamma \approx0.12$~$\mu$eV,
corresponding to an inhomogeneous dephasing time $T_{2}^{\ast}=\sqrt{2}/\Gamma\approx8$~ns.
The exchange energy $J_{\text{ex}}\approx-1$~$\mu$eV is determined with the Hund-Mulliken method~\cite{LossPRB2070},
consistent with experimental values~\cite{Marcus_T2}.

Figure~\ref{fig1}~(b) shows the FID decoherence for several nuclear spin initial states $|{\mathcal I}\rangle$
randomly chosen from the thermal ensemble. The suppression of the decoherence with decreasing the Overhauser field
mismatch ($\Delta_{\mathcal I}$) is evident. The dependence of the decoherence time on the Overhauser field mismatch
${\mathbb T}_{2}^{\mathcal I}\propto\left(\Delta_{\mathcal I}/E_{\text{S-T}}\right)^{-1/2}$ is verified
in Fig.~\ref{fig1}~(c). Note that for vanishing mismatch $\Delta_{\mathcal I}\rightarrow 0$, the energy
cost ${\mathbb Z}_k$ vanishes in
the leading order of ${Z}_k$ and the second-order correction ${\mathbb Z}_{k}={Z}_{k}^2/J_{\text{ex}}$, resulting
in a saturation of the decoherence time at a large value. In ensemble dynamics, the signal is to be averaged by
${\mathbb L}({\tau})=\sum_{\mathcal I} P_{\mathcal I} {\mathbb L}_{\mathcal I}({\tau})$.
The inhomogeneous broadening of the Overhauser field results in an effective
dephasing $\sum_{\mathcal I} P_{\mathcal I}e^{-iE_{\text{S-T}}{\tau}}$ with a nanosecond timescale~\cite{Coish_ST,Taylor:2006},
much faster than the entanglement-induced decoherence ${\mathbb L}_{\mathcal I}({\tau})$ in single-sample dynamics.
So below we study the ensemble-averaged coherence in spin-echo configurations where
the inhomogeneous broadening effect is eliminated.

\begin{figure}[ptb]
\includegraphics[width=\columnwidth]{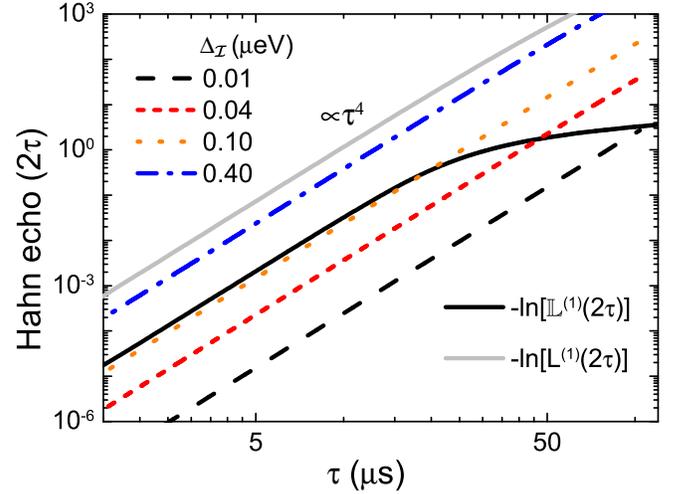}
\caption{(Color online) Hahn echo signal ($-\ln [{\mathbb L}^{(1)}_{{\mathcal I}}(2\tau)]$)
for various nuclear spin configuration $|{\mathcal I}\rangle$ as indicated by $\Delta_{\mathcal I}$.
The solid black line shows the ensemble-averaged coherence ($-\ln [{\mathbb L}^{(1)}(2\tau)]$),
compared to the echo signal ($-\ln[L^{(1)}(2\tau)]$) of a single spin in a QD of the same size
(solid gray line).}
\label{fig2}
\end{figure}

In the single-pulse Hahn echo ($\tau$-$\pi$-$\tau$-echo), the S-T coherence at the echo time $2\tau$
for a nuclear spin state $|{\mathcal I}\rangle$ is
\begin{equation}
{\mathbb L}_{\mathcal I}^{(1)}(2\tau)=\prod_k\left|\left\langle\downarrow\left|
 \left(\hat{U}^{(1),-}_k\right)^{\dag}\hat{U}^{(1),+}_k\right|\downarrow\right\rangle\right|,
 \label{Hahn}
\end{equation}
where $\hat{U}^{(1),\pm}_k\equiv  e^{-i{\boldsymbol \chi}_k^{\mp}\cdot\hat{\mathbf s}_k\tau}e^{-i{\boldsymbol \chi}_k^{\pm}\cdot\hat{\mathbf s}_k\tau}$.
The short-time  behavior (for $\tau\ll {\mathbb Z}_k^{-1}$)
is ${\mathbb L}_{\mathcal I}^{(1)}(2\tau)\approx\exp\left(-2\sum_{k}B_{k}^{2}{\mathbb Z}_{k}^{2}\tau^{4}\right)\equiv e^{-\left(2\tau/{\mathbb T}_H^{\mathcal I}\right)^4}$
with the single-sample decoherence time
${\mathbb T}_{H}^{\mathcal I}=\left(\sum_{k}B_{k}^{2}{\mathbb Z}_{k}^{2}/8\right)^{-1/4}=\sqrt{2}{\mathbb T}_{2}^{\mathcal I}$.
The ensemble dynamics is studied by averaging over a large number of samples from a Gaussian distribution
of the Overhauser field. With the approximation $E_{\text{S-T}}\approx J_{\text{ex}}$, the ensemble-averaged
result is analytically obtained for $\tau\ll {\mathbb Z}_k^{-1}$,
\begin{equation}
{\mathbb L}^{(1)}(2\tau)\approx\left[ 1+\left(2\tau/{\mathbb T}_{H}\right)^{4}\right]^{-1/2},
\end{equation}
with a power-law decay profile, where the ensemble decoherence time
${\mathbb T}_H=\left. {\mathbb T}^{\mathcal I}_H\right|_{\Delta_{\mathcal I}=\sqrt{2}\Gamma}$,
i.e., the decoherence time for a nuclear spin configuration with the Overhauser field
mismatch equal to $\sqrt{2}$ times the standard variance.
Such a transition from an exponential decay in single-sample dynamics to a
power-law decay in ensemble dynamics is shown in Fig.~\ref{fig2}.
In contrast, the echo signal of a single electron spin in a QD is unchanged by
ensemble averaging, i.e., ${L}^{(1)}={L}^{(1)}_{\mathcal I}$.

\begin{figure}[ptb]
\includegraphics[width=\columnwidth]{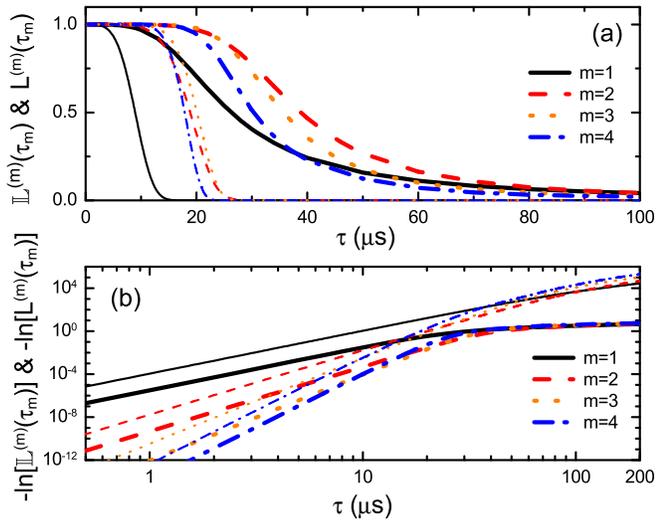}
\caption{(Color online) (a) Ensemble-averaged coherence under concatenated control,
for the S-T decoherence in two coupled dots (${\mathbb L}^{(m)}(\tau_{m})$, as thick lines),
and for the single spin decoherence in one dot (${L}^{(m)}(\tau_{m})$, as thin lines), where
$m$ indicates the concatenation level. (b) Logarithmic plot of (a).
}
\label{fig3}
\end{figure}

We now study the S-T decoherence under concatenated control which is designed to
preserve the coherence~\cite{Yao_DecoherenceControl,Liu_Decoherence,Lidar_CDD}.
The coherence preserved after the $m$th order concatenated pulse sequence
${\mathbb L}^{(m)}_{\mathcal I}(\tau_m)$ is obtained by substituting
$\hat{U}^{(m),\pm}_k$ for $\hat{U}^{(1),\pm}_k$ in Eq.~(\ref{Hahn}),
where $\tau_m\equiv 2^m\tau$, and $\hat{U}^{(m),\pm}_k$ is recursively defined as $\hat{U}^{(m-1),\mp}_k\hat{U}^{(m-1),\pm}_k$ for $m>1$.
The short-time profile (for $\tau\ll {\mathbb Z}_k^{-1}$) is
${\mathbb L}^{(m)}_{\mathcal I}(\tau_m)\approx e^{-\left(\tau_m/{\mathbb T}^{\mathcal I}_{2,m}\right)^{2m+2}},
$
with the decoherence time ${\mathbb T}^{\mathcal I}_{2,m}\sim 2^{\frac{m(m+3)}{2(m+1)}}\left[{Z}_kB_k^{m}\left(\Delta_{\mathcal I}/E_{\text{S-T}}\right)\right]^{-\frac{1}{m+1}}.$
Again, the ensemble average leads to a power-law decay
\begin{equation}
{\mathbb L}^{(m)}(\tau_{m})=\left[1+({\tau_{m}}/{{\mathbb T}_{2,m}})^{2m+2}\right]^{-1/2},\label{Ln}
\end{equation}
with ${\mathbb T}_{2,m}=\left.{\mathbb T}^{\mathcal I}_{2,m}\right|_{\Delta_{\mathcal I}=\sqrt{2}\Gamma}$.
In contrast, for a single electron spin, the ensemble-averaging has negligible effect,
i.e., ${L}^{(m)}(\tau_{m})={L}^{(m)}_{\mathcal I}(\tau_{m})\approx e^{-(\tau_{m}/{T}_{2,m})^{2m+2}},$
where the decoherence time ${T}_{2,m}$ is shorter than the S-T decoherence time ${\mathbb T}_{2,m}$ by
a factor $\sim\left(J_{\text{ex}}/\Gamma\right)^{1/(m+1)}$.
Fig.~\ref{fig3} compares the S-T decoherence to the single spin decoherence,
showing the suppression of the decoherence and the crossover to a power-law decay
due to the coupling between the electron spins.

In conclusion, the exchange interaction between two electron spins in double QDs
modifies the nuclear spin bath dynamics through renormalizing the pair-flip excitation energy.
As the renormalized excitation energy varies with the Overhauser field mismatch between the
two dots, the nuclear spin dynamics in one dot becomes dependent on the nuclear spin state in
the other dot, regardless of nonexistence of inter-dot nuclear-spin interaction in the
considered situation. Consequently, the singlet-triplet decoherence due to the electron-nuclear entanglement
depends on choice of the nuclear spin configuration from an ensemble, leading to a power-law
decay of ensemble-averaged coherence, in contrast with the exponential decoherence of a single
electron spin which is insensitive to sampling of the nuclear spin ensemble. The dependence of
the S-T decoherence on the Overhauser field mismatch may be observed by tuning the mismatch with
an inhomogeneous external field. The exchange interaction also enhances the S-T decoherence time
by suppressing the fluctuation in the nuclear spin bath.

This work was supported by Hong Kong RGC Project 2160285.


\begin{thebibliography}{26}
\bibitem[{not()}]{note_ST_mediated}
\bibinfo{note}{In principle, the virtual flip-flops ($\left\vert
  S\right\rangle$ or $|T_0\rangle \leftrightarrow \left\vert
  T_{\pm}\right\rangle$) via the off-diagonal hyperfine interaction mediate the
  indirect interaction between nuclear spins (being or not in the same dot),
  with strength $\sim a_{j,n} a_{j',n'} J_{\text{ex}}/\Omega^2_{e}$, which,
  however, is negligible as $\Omega_{e}\ggg J_{\text{ex}}, a_{j,n}$.}
\bibitem[{Not()}]{Note_on_single}
\bibinfo{note}{For two uncorrelated electrons in double QDs
  ($J_{\text{ex}}=0$), the $T_2$ pure dephasing of the two independent electron
  spins is a $T_1$ longitudinal relaxation in the S-T basis.}
\end{thebibliography}

\end{document}